\documentclass{emulateapj}

\slugcomment {Accepted by the Astronomical Journal}

\begin{document}

\title{The SN 393 -- SNR RX J1713.7-3946 (G347.3-0.5) Connection }

\author{Robert A.\ Fesen{\altaffilmark{1}}, Richard Kremer{\altaffilmark{2}},
        Daniel Patnaude{\altaffilmark{3}}, \& Dan Milisavljevic{\altaffilmark{1}}}

\altaffiltext{1} {6127 Wilder Lab, Department of Physics \& Astronomy, Dartmouth
  College, Hanover, NH 03755  }

\altaffiltext{2} {6107 Carson Hall, Department of History,  Dartmouth
  College, Hanover, NH 03755  }

\altaffiltext{3} {Smithsonian Astrophysical Observatory, Cambridge, MA 02138 }

\begin{abstract}

Although the connection of the Chinese `guest' star of 393 AD with the Galactic
supernova remnant RX J1713.7-3946 (G347.3-0.5) made by Wang et al.\ in 1997 is
consistent with the remnant's relatively young properties and the guest star's
projected position within the `tail' of the constellation Scorpius, there are
difficulties with such an association.  The brief Chinese texts concerning the
393 AD guest star make no comment about its apparent brightness stating only
that it disappeared after 8 months.  However, at the remnant's current
estimated $1 - 1.3$ kpc distance and $A_{\rm V} \simeq 3$, its supernova should
have been a visually bright object at maximum light ($-3.5$ to $-5.0$ mag) if
$M_{\rm V}$ = $-17$ to $-18$ and would have remained visible for over a year.
The peak brightness $\simeq$ 0 magnitude adopted by Wang et al.\ and others
would require the RX J1713.7-3946 supernova to have been a very subluminous
supernova event similar to or fainter than CCSNe like SN~2005cs.  We also note
problems connecting SN 393 with a European record in which the Roman poet
Claudian describes a visually brilliant star in the heavens around 393 AD that
could be readily seen even in midday.  Although several authors have suggested
this account may be a reference to the Chinese supernova of 393, Scorpius would
not be visible near midday in March when the Chinese first reported the 393
guest star.  We review both the Chinese and Roman accounts and calculate
probable visual brightnesses for a range of supernova subtypes and conclude
that neither the Chinese nor the Roman descriptions are easily reconciled with
an expected RX J1713.7-3946 supernova brightness and duration.

\end{abstract}


\section{Introduction}

Since the middle of the 19th century, astronomers have known about pre-Tychonic
bright new or `temporary' stars recorded in ancient Asian, Arabic, and European
texts chiefly through the works of \citet{Biot1846}, \citet{Chambers1867},
\citet{Williams1871}, \citet{Humboldt1871}, \citet{Clerke1890}, and
\citet{Lundmark21}. With the discovery of supernovae (SNe) early in the 20th
century \citep{BZ34}, extensive searches for historic sightings of Galactic SNe
were made by \citet{Kanda35}, \citet{Hsi55}, and
\citet{HoPengYoke62,HoPengYoke66},  with \citet{CS76,CS77} and \citet{SG02}
presenting detailed analyses of the most likely historic SN sightings along
with relevant cultural and background material.

Only a handful of celestial objects reported between the years 1 and 1000 AD 
have descriptions or durations indicating a likely Galactic SN event
(see \citealt{SG02} for details).  Among these are three Chinese `guest'
stars reported during the second half of the 4th century remarkably
separated by less than 25 years; namely, the new stars of 369, 386, and 393 AD.
Of these the most likely SN event is the one seen in 393 due to its
nearly eight month long period of visibility.  

The Chinese description of the 393 star is contained in just two brief and
nearly identical records from the Jin dynasty (the {\it Songshu} and the {\it
Jinshu}; \citealt{CS77,Wang97,Xu2000,SG02}).  The translated texts state that
during the second month of the 18th year (27 Feb -- 28 Mar of 393 AD) a guest
star appeared within {\it{Wei}} (the tail of Scorpius) and lasted until the
ninth lunar month (22 Oct -- 19 Nov) when it was extinguished [or disappeared].

\citet{SG02} note that the use of the term `{\it{zhong}}' used in the text
meaning `within' is unique among celestial descriptions recorded during the
Jin dynasty and hence strongly implied that the 393 star appeared within the
bowl-like boundaries of Scorpius' tail.  The Galactic plane virtually bisects
the well-defined Chinese asterism {\it{Wei}} making up the tail of Scorpius
(see Fig.\ 2 in \citealt{Wang97}), consistent with a nova or supernova occurring
in or close to the Galactic plane. 

\citet{Psk72} argued that the 393 star was likely a recurrent nova since
Chinese records also reported a star with {\it Wei} in 1600 AD.
A nova interpretation was also suggested by \citet{vdb78}  due to the
lack of an optically bright supernova remnant in that region. However, the
star's eight month long period of visibility, ending only when it began setting
around sunset, has been viewed as strong evidence for it being a Galactic supernova.

Identifying the remnant of a historic SN is often difficult and the guest star
of 393 is no exception, with nearly a dozen Galactic supernova remnants located
within the tail of Scorpius \citep{Green09}.  The remnants of G348.5+0.1 (CTB
37A) and G348.7+0.3 (CTB 37B) were initially seen as possible SNR candidates to
the 393 guest star due to their small angular sizes of $15'$ and $17'$,
respectively \citep{CS77,SG02}. However, these remnants lie $\simeq$ 10 kpc
away \citep{Rey00,Aharonian08,Nak09} and such large distances near the Galactic
center likely imply considerable interstellar extinction decreasing the chance
of an associated visually bright guest star. The same is true for the
apparently very young SNR G350.1-3.0 whose distance is only $\sim$3.4 kpc but
lies behind an estimated $\simeq$ 20 mag of visual extinction \citep{Gae08}.

In 1996 \citet{Pfeffermann96} announced the {\sl ROSAT} discovery of the
Galactic remnant RX J1713.7-3946 (G347.3-0.5) in Scorpius.  The remnant's
location near the SN 393 reported position along with Pfeffermann and
Aschenbach's estimated remnant distance of $1.1$ kpc and 2100 yr age led
\citet{Wang97} to suggest it as the likely remnant of SN 393.  Although
\citet{SG02} and \citet{Nick10} have argued for the remnant's
distance of $6 \pm1$ kpc based on possible associations with nearby molecular
clouds \citep{Slane99} which would rule out its association with SN 393, more recent
distance estimates of RX J1713.7-3946 firmly place it between
$0.9 - 1.7$ kpc, and most recent papers on RX J1713.7-3946 cite 
the \citet{Wang97} proposed SN 393 connection. 

We note here, however, that a distance less than 2 kpc for RX J1713.7-3946
raises problems with the expected supernova maximum apparent brightness and
durations which appear in conflict with the Chinese records.  Below we discuss
how neither the Chinese record of the guest star nor a long known but rarely
cited European account of a bright star reported about that same year are
consistent with RX J1713.7-3946 as the remnant of the probable Chinese
supernova of 393 AD.

\section{The RX J1713.7-3946 Remnant}

The remnant RX J1713.7-3946 exhibits several properties suggesting a relatively
young age, probably less than a few thousand years and thus potentially
consistent with a SN event around 393 AD. Its X-ray emission is that of a
featureless nonthermal continuum consistent with a shock velocity of several
1000 km s$^{-1}$ \citep{Koyama97,Slane99,Uch03,Uch07,Fukui03,CC04}.  While the
shock's velocity is probably high, it is likely less than 4500 km s$^{-1}$
based on a limit of the angular displacement of the remnant's outer X-ray
emission over a 6 year period \citep{Uch07}. \citet{Ellison10} found velocities
$\sim$ 3000 km s$^{-1}$ were required in order to model the shock kinematics in
an evacuated cavity at an assumed age of 1600 years.

RX J1713.7-3946 is also one of only a few Galactic SNRs which exhibit
gamma-rays with energies up to 100 TeV.  In this respect it resembles several
other young Galactic SNRs including RCW~86 (SN 185) and the SN 1006 remnant.
Observations show a close correlation between the remnant's X-ray and gamma-ray
emissions suggesting a causal connection between the processes generating both
types of emissions \citep{Aharonian07}. Since high shock velocities are
required to generate a significant nonthermal X-ray and gamma-ray flux
\citep[see e.g.][]{zirakashvili07}, the presence of an X-ray synchrotron
emission and coincident gamma-ray emission in RX J1713.7-3946 favor a distance
of $\sim$ 1 kpc, which would rule out the age of 20,000--40,000 yr which is
based on the much larger distance estimate of 6 kpc \citep{Slane99}.

The presence of a compact X-ray source 1WGA J1713.4-3949 at a projected
location near the remnant's center with a similar X-ray derived $N_{H}$
column density to the remnant's central regions \citep{CC04} implies it is 
associated with the remnant thereby indicating the remnant is
from a core-collapse SN (CCSN).  The remnant's $65' \times 55'$ angular size 
and recent distance estimates $\sim 1 - 1.3$ kpc \citep{Fukui03,Moriguchi05} when
combined with a high-velocity shock suggests 
a relatively low ambient ISM density ($\sim0.01$ cm$^{-3}$) like that expected
in a stellar wind driven cavity generated by a high-mass progenitor which 
exploded as a Type II/Ib,c SN \citep{CC04}.

\bigskip

\section{Problems with a SN 393 - RX J1713.7-3946 Connection}

Of the currently known SNRs located near the Chinese reported position of SN
393, RX J1713.7-3946 would seem a good candidate remnant for the presumed
supernova of 393.  It is a relatively young SNR and lies within the tail
of Scorpius consistent with the Chinese report about the location of the 393
guest star. However, there are difficulties with this SN--SNR connection.
Below we discuss these difficulties along with a European report of a bright daytime
star around 393 AD but which is unlikely to be a sighting of the Chinese 393 SN.

\subsection{Expected Brightness of a RX J1713.7-3946 SN}

As shown in Table 1, recent distance estimates for RX J1713.7-3946 range
between  $0.9 - 1.7$ kpc
(\citealt{Fukui03,Koo03,Koo04,Aharonian04,CC04,Moriguchi05}) with a
concentration of recent values around $1 - 1.3$ kpc. This would place an RX
J1713.7-3946 supernova closer than any historic Galactic SN recorded during the
past two millennia.  A distance of just $1 - 1.3$ kpc also means that its SN
should have been visually bright if: a) the optical extinction to it is fairly
low, and b) it was not an unusually faint event.

X-ray derived $N_{\rm H}$ column densities for the remnant's central X-ray
emitting regions range from $ 4 - 8 \times 10^{21}$ cm$^{-2}$ depending on the blackbody
or power law model adopted \citep{CC04}.  Adopting $A_{\rm V} = N_{H}/1.8
\times 10^{21}$ cm$^{-2}$ for a typical gas-to-dust ratio
\citep{Bohlin78,PS95,KM96}, yields $A_{\rm V}$ values around 2.2 -- 4.4
mag. In the following discussions, we will adopt an $N_{H} = 5 \times 10^{21}$
cm$^{-2}$ consistent for the remnant's central compact X-ray source 1WGA
J1713.4-3949 modeled by a two-component blackbody \citep{CC04}, which translates to 
$A_{\rm V} \simeq 2.8$ or roughly 3 magnitudes of extinction.

As shown in Tables 1 and 2, if the RX J1713.7-3946 SN had an $M_{\rm V}$ of $-17$ to
$-18$ in line with the typical absolute visual magnitudes for core-collapse
SNe~II or SNe~Ib,c \citep{Richardson02,Richardson06,Drout11}, then the RX
J1713.7-3946 supernova should have been a visually brilliant object. For example, at a
distance between 1.0 and 1.3 kpc and an $A_{\rm V}$ = 3.0, a SN would have been
$m_{\rm v}$ = $-3.5$ to $-5.0$ for $M_{\rm V}$ = $-17$ to $-18$.  In fact, it
would have been a bright guest star almost regardless of the specific SN
subtype, i.e., Type Ia, II, Ib,c, or IIn unless it was a
very subluminous event.

Pushing the numbers toward fainter apparent magnitudes limits by adopting the
largest distance recently estimated for RX J1713.7-3946, namely 1.7 kpc, and an
$A_{\rm V}$ = 4 ($N_{H}  = 7 \times 10^{21}$ cm$^{-2}$; \citealt{CC04}), and a
$M_{\rm V}$ of $-16.0$ equal to the faintest Type Ib,c SNe \citep{Drout11}, the
supernova would still have been quite bright with $m_{\rm v} \approx -1$ and
should have remained visible well into 394 AD.  Smaller distances of around 1
kpc as suggested by \citet{Fukui03} and \citet{Moriguchi05} but keeping $A_{\rm
V}$ = 4.0 leads to apparent magnitudes around $-2$, or about as bright as
Jupiter.

To place these brightness estimates in context, we list in Table 1 the
reported peak visual magnitudes (to the nearest half magnitude where possible)
of all Galactic SN over the last two millennia based on ancient records
\citep{SG02}. The listed peak magnitudes, especially for the older SNe, are
often uncertain due to the sometimes fragmentary nature of the existing records and
possible observer error or exaggeration.  For example, whereas the estimated
distance and visual extinction to the SN 1006 remnant implies a maximum
apparent visual brightness around $-7.5$ \citep{Winkler03}, analyses of
reported descriptions suggest values 2 mag higher or lower
\citep{Stephenson10}.  This plus uncertainty in remnant distances has led to
somewhat different values listed in Table 1 from other authors
(e.g., \citealt{Schaefer96}).

Despite these caveats, one finds that the peak magnitude estimates from ancient
descriptions is, in most cases, in rough agreement with visual magnitude
estimates based upon the likely SN event. We have identified core-collapse
supernova events in cases where a compact central object is present in the
likely associated SNR. Listed in Table~\ref{tab:hist} are the current estimates
for remnant distances and line of sight visual extinctions. Besides the 393 SN,
the worst agreement between the predicted and reported brightness is perhaps
that of SN 1181 with the remnant 3C58 about which there has been considerable
debate as to that remnant's age and distance estimates
(see \citealt{Fesen08} and \citealt{Kothes10} and references therein).

\subsection{Estimates on the Apparent Brightness of SN 393}

As noted above, the Chinese records include no comment about the star's
brightness. This leaves one to wonder why they didn't include a comment about
brightness especially if SN 393 is related to RX J1713.7-3946 it might have
rivaled Jupiter or even Venus at its brightest. One possible solution might
simply lie in the extreme brevity of the existing Chinese records.  The
description is about as short as one could write a record concerning the
appearance of a guest star.

In considering this peak brightness issue, it is important to note that the
records that have survived from this period in China are condensed summaries of
the Jin dynasty history written by imperial scholars many decades and even
centuries after the actual events.  Moreover, trying to interpret the meaning
of the lack of a note about the star's brightness is made more difficult given
the fact that no mention is made in the existing records on the brightnesses of
either of the other two possible guest stars of 369 and 386.

In the absence of any indication in the Chinese records regarding the
brightness for the 393 guest star, \citet{CS77} estimated a peak visual
magnitude $\simeq 0$. They arrived at this value from the lack of any
remark about any extraordinary brightness and a consideration of atmospheric
extinction since observers at Nanjing could view the guest star no more than
20 degrees above the horizon.

In like fashion, \citet{Wang97} estimated it could have been no brighter than
$-2$ magnitude, reasoning that had the guest star been brighter than this the
Chinese astronomers would have compared it to the planets Saturn, Mars and
Jupiter. All three planets were, in fact, visible in the morning sky in the
early Spring of 393 AD with apparent visual magnitudes of $+$0.27, $+$0.25, and
$-$2.4 respectively, and located within 45 degrees of the reported guest star
in Scorpius.

\citet{Wang97} further argued that if its peak magnitude had been fainter than
0 mag then it would have not been visible to the naked eye for as long as 8
months.  Adopting a distance of 1.1 kpc, an $A_{\rm V}$ = 2.0, and a peak
visual magnitude of $-2$ to $0$, they estimated a RX J1713.7-3946 supernova to
have been $M_{\rm V}$ = $-12$ to $-14$. 
However, this would mean SN 393 was an unusually low-luminosity supernova event
similar to SN 1987A, 1999br, and 2005cs with $M_{\rm V} >  -15$
\citep{Richardson02,Pastorello2004,Pastorello2006}.
On the other hand, if the RX J1713.7-3946 supernova had
a more common $M_{\rm V}$ for core-collapse SNe of $-17$ to $-18$, it then
should have been easily visible into early 394 AD. Yet there is no mention of
this in the existing Chinese records.  So, if the 393 event only reached $-2$ to $0$
mag and was connected to RX J1713.7-3946, then at the distance and visual
extinction assumed by \citet{Wang97} it must have been a very subluminous event ($M_{\rm V} > -15$)
similar to SN II/Ib,c events occupying the faint end of CCSNe
\citep{Richardson02}.

\subsection{Reported Duration of SN 393}

In March 393 when the Chinese reported the first sighting of the 393 guest
star, astronomers in Nanjing (latitude $+32\arcdeg$) would have been able to
see the stars comprising Scorpius' tail rise above the horizon around 1 am
local time and reach culmination 22 degrees above the southern horizon in the
morning by 5 am, roughly a half hour before the beginning of astronomical
twilight.  If the guest star was as brilliant as we estimate if it were the RX
J1713.7-3946 supernova ($-4.5$ mag), it would have been easy to follow from
night into twilight and then into daytime. 

Such a guest star in Scorpius would remain visible at night for the next
several months right up until the time it would set in evening twilight around
mid-September. A star about as bright or brighter than the nine stars
comprising the Scorpius's tail asterism, all but one of which are fainter than
2nd magnitude, might stay visible a little longer. Thus it might have still been
visible during early twilight a week or two longer and hence possibly into
early October.  This would still fall short of the Chinese records which states
it lasted until the ninth lunar month, namely 22 October through 19 November.
Of course the description of  `until' the ninth month could mean that its
visibility was only up to the 9th month and not during it.

\citet{SG02} interpret the duration of the guest star
`until' the Chinese 9th lunar month (22 October -- 19 November) as meaning the
star remained visible into the the ninth month.  However, on October 22 it
would have set just some 15 minutes after sunset and to be
visible under these circumstances the star would have required a relatively
bright object, thereby implying a brilliant object months earlier at
maximum light.

In an attempt to resolve this dilemma, Stephenson and Green
propose a recording error of one month regarding the object's disappearance
(the 8th instead of the 9th month), allowing the object to set well after
sunset in a dark sky.  Considering the visibility of Scorpius' asterism {\it
Wei} in late September and early October 393 and allowing for atmospheric
extinction, \citet{CS77} estimate a maximum apparent magnitude around $0$ mag or
perhaps a bit brighter, noting that had the star been much brighter than this
the Chinese likely would have included a comment on its brightness.

In Table~\ref{tab:lc}, we list the expected brightnesses eight months (day 240)
after maximum light for several different CCSN subtypes assuming a distance and
$A_{\rm V}$ equal to that of  RX J1713.7-3946. The table shows that in all
cases the supernova1G would have been brighter than $\theta$ Sco ($m_{\rm v}$ =
1.87), the brightest star in the tail of Scorpius. Thus, if RX J1713.7-3946 is
a CCSN and was the guest star seen in 393, then its expected brightness between
0 and +1.5 mag would make it possible for it to stay visible a little longer
but perhaps not past the middle of October as the Chinese reported.

A more serious problem with this scenario is that there is no mention of the
star being recovered early in 394 AD when it would have again become visible
from behind the Sun.  Given average luminosity decline times commonly seen for
core-collapse supernovae, a RX J1713.7-3946 supernova should have been easily
visible to observers with an apparent brightness between 1st and 3rd mag,
comparable to the stars in Scorpius's tail (see Table~\ref{tab:lc}). Although
the very brief Chinese record should not be interpreted as complete, it is
unusual that there were no further reports of its continued presence,
especially since reports of other guest stars returning from behind the Sun
exist, such as SN 185.

However, in light of the considerable spread in the decline rates of SN Type
Ib,c events, the possibility exists that the supernova faded below widespread
notice when it came from around the Sun three months later.  Some subluminous
events exhibit a steepening of their light curve at times beyond 120 days,
diminishing in visual brightness fairly rapidly.  For instance, had a RX
J1713.7-3946 supernova followed the light curves of the SN~2005cs or SN~2009md
(\citealt{Pastorello2009,Fraser2011}), it would have faded $>$ 5 mag one year
past maximum, and hence possibly would have been missed.

In closing, we note that if SN 393 had instead been a Type Ia event and
unrelated to the CCSN remnant RX J1713.7-3946, similar brightness issues at
late times would apply.  That is, at day 240 a Type Ia guest star would appear
$\simeq$ 6 magnitudes fainter than at maximum light \citep{Lei91}. A peak
brightness of 0 mag estimated by \citet{SG02} would mean the guest star would
approach the naked eye visibility limit of 6th mag some  months after maximum.
This would make the star even more difficult to view in early
October 393 since it would be less than 5 degrees above the western horizon at
the end of twilight and thus subject to significant atmospheric attenuation.

\section{A Possible European Sighting of SN 393 }

Relevant to the apparent brightness of the SN 393 guest star, there is an  
European text written around 398 AD by the Roman poet
Claudian describing a bright star about which he said was plainly visible
even in midday a few years earlier.  Claudian viewed this star as an omen
of Honorius being made emperor in 393\footnote{Honorius was declared emperor
{\it Augustus} at the age of nine in 393 by his father Theodosius I. With the
death of their father in January 395, he and his brother Arcadius divided up
the empire, with Honorius becoming the Western Roman emperor.} thereby implying
that it occurred around 393 AD.

The possibility of a connection between the bright star described in the
Claudian poem and the Chinese guest star of 393 AD has been made by several
authors including \cite{Dreyer1913}, \citet{Stothers77}, \citet{Barrett78},
\citet{CS82}, \citet{Clark84}, and \citet{Ramsey2006}.  However, no mention of
this reference is found in the most recent astronomical reviews of ancient
guest star observations including discussions directly regarding the suspected
SN of 393 \citep{CS75,CS77,Wang97,SG02,GS03,Wang06}.

Interestingly, this Roman record has long been known, going back some 440 years
to the time of Tycho Brahe.  A year after he sighted his supernova of 1572,
Tycho learned about the Claudian text through a letter from Paul Hainzel, a
longtime friend and then mayor of Augsburg, to the humanist Hieronymous Wolf in
which Hainzel mentions the Claudian text about a bright new star in the sky,
much like the 1572 star \citep{Tycho1602}.  Tycho never reached a definitive
conclusion about the meaning of the Claudian poem, i.e., whether it was a
description of a comet, a daytime sighting of Venus, or something else, but it
was obvious, he concluded, that new stars like the one he saw in 1572 sometimes
appear in the heavens \citep{Dreyer1913}. 

The relevant passages about the new star appear in Claudian's `The Fourth
Consulship of Honorius'.  Since to our knowledge this text has never been
presented in the astronomical literature, we reproduce it here. According to
the English translation of the Latin by \citet{Platnauer22}, the pertinent
passages read:

\begin{quote}

Never was the encouragement of the gods more sure, never did heaven attend with
more favouring omens.  Black tempest had shrouded the light in darkness and the
south wind gathered thick rain-clouds, when of a sudden, so soon as the
soldiers had borne thee aloft with customary shout, Phoebus scattered the
clouds and at the same moment was given to thee the sceptre, to the world
light. Bosporus, freed from clouds, permits a sight of Chalcedon on the farther
shore; nor is it only the vicinity of Byzantium that is bathed in brightness;
the clouds are driven back and all Thrace is cleared; Pangaeus shows afar and
lake Maeotis makes quiver the rays he rarely sees.  'Tis not Boreas nor yet
Phoebus' warmer breath that has put the mists to flight. 

That light was an emperor's star. A prophetic radiance was over all things, and
with thy brightness Nature laughed. Even at midday did a wondering people gaze
upon a bold star ('twas clear to behold) no dulled nor stunted beams but
bright as Bo\"otes' nightly lamp.  At a strange hour its brilliance lit up the
sky and its fires could be clearly seen though the moon lay hid. May be it was
the Queen mother's star or the return of thy grandsire's now become a god, or
may be the generous sun agreed to share the heavens with all the stars that
hasted to behold thee. 

The meaning of those signs is now unmistakable.  Clear was the prophecy of
Ascanius' coming power when an aureole crowned his locks, yet harmed them not,
and when the fires of fate encircled his head and played about his temples.
Thy future the very fires of heaven foretell. 

\end{quote}

Although a similar translation is found in \citet{Barr81}, both these authors
have taken some poetic license with the text. A more
literal translation of the key sentence and a slight modification of that given
by \cite{Ramsey2006} is: ``Even at midday, the marveling populace beheld a bold
and unmistakable star, which was not faint with dimmed ray, but as bright as
Bo\"otes is at night. It shone forth, a guest in fiery regions at a strange
hour, and it could be recognized although the moon lay hidden.''

Although Claudian's description concerning a star being ``bold'' (audax) and
visible even at midday suggesting a very bright object would 
seem consistent with our estimated peak magnitude around $-4.5$ for a RX
J1713.7-3946 supernova, it appears unlikely that it is a reference to the
Chinese guest star of 393.  The constellation Scorpius would set by 9 am in the
morning of early March 393 and hence no star located in its tail would be
visible near midday.

One possibility is that Claudian's star is a sighting of a brilliant $-4.5$ or
brighter SN in late 392 when Scorpius would be near conjuction with the Sun.
However, in that case the Chinese then should have reported it in the early
morning by mid-January when Scorpius rises an hour before morning twilight.
Additionally, there is no Chinese record concerning a daytime star in 392.   

Interestingly Venus, which can be seen during daylight when brighter than
$\approx -3.5$ magnitude \citep{Weaver1947}, was near its maximum brilliance of
$-4.6$ mag and near the meridian at 9 am on 23 January 393 when Honorius was
declared emperor (Augustus) by his father Theodosius. At noon on that day,
Venus would still have been visible some 25 degrees above the southwestern
skies. The waning crescent Moon would have already set by midday and might
explain Claudian's reference to a star's visibility ``though the moon lay
hid''.  So maybe a daytime sighting of Venus is what Claudian was referring to
in his poem of adoration to emperor Honorius.

\section{Conclusions}

Identifying the remnants of historic SNe is often difficult and this is
especially true in the case of the 393 guest star.  Given the brief description
of the guest star in the Chinese records and the nearly dozen Galactic
supernova remnants currently known within the tail of Scorpius \citep{Green09},
doubts about any SN 393 -- SNR association will likely persist.

A connection between the ancient guest star and suspected supernova of 393 and
the X-ray bright supernova remnant RX J1713.7-3946 has been proposed by
\citet{Wang97}.  While this connection has been often cited in the literature
on the RX J1713.7-3946 remnant and is in line with some of its relatively
youthful properties, the \citet{Wang97} estimated $M_{\rm V}$ values between
%%%%%%%%%%%%%%%%%%%%%%%%%%%%%%%%%%%%%%%%%%%
%%%% Change the end of sentence to read:
$-12$ and $-14$ imply a very subluminous core-collapse SN event.  
%%%%%%%%%%%%%%%%%%%%%%%%%%%%%%%%%%%%%%%%%%%

In this paper, we reviewed both the Chinese and Roman accounts and calculated
probable visual brightnesses for a range of supernova subtype. We conclude that
neither the Chinese nor the Roman descriptions are easily reconciled with an
expected RX J1713.7-3946 supernova brightness and duration.  We further note
that if RX J1713.7-3946 were the SN 393 remnant, it would then rank as having
been the nearest of all the known historic Galactic supernovae during the last
2000 years.  It's relatively small distance of around 1 kpc plus a moderate
amount of optical extinction also means its supernova would have likely have
been a visually brilliant object, certainly as bright as Jupiter and maybe as
bright as Venus.

Although a connection between SN 393 and RX J1713.7-3946, or for that matter
any other young SNR lying within Scorpius' tail, will likely remain uncertain
due to limitations of the ancient records, such an association does not appear
consistent with the available historical records.  It is hoped that future
studies of the RX J1713.7-3946 remnant which provide better estimates as to its
age may help resolve the question of the remnant of the suspected SN of 393 AD.

\acknowledgements

\newpage

\begin{deluxetable}{rccccccc}
\tabletypesize{\scriptsize}
\tablecolumns{8}
\tablecaption{Observed Galactic Supernovae Over the Last Two Millennia in Order of Increasing Distance }  
\tablewidth{0pt}
\tablehead{ \colhead{SN} & \colhead{Reported\tablenotemark{a}}  & \colhead{Duration} &   \colhead{ Confirmed or Proposed} & \colhead{SN\tablenotemark{b}} 
&  \colhead{SNR Distance } &  \colhead{$A_{\rm V}$}   &  \colhead{Expected\tablenotemark{c}} \\ 
\colhead{} &  \colhead{$m_{\rm v}^{\rm max}$}     & \colhead{(months)} & \colhead{SN Remnant} & \colhead{Type}  & \colhead{(kpc)}  & \colhead{(mag)}  &
\colhead{$m_{\rm v}^{\rm max}$}  } 
\startdata
 393    &   $\sim0~$   &   $8$    & RX J1713.7-3946 & CCSN  &   0.9--1.7   & 3.0   & $-3.8$ to $-5.2$  \\
1054    &   $-5.0$     &   $21$   &  Crab Nebula    & CCSN  &   1.8--2.0   & 1.6   & $-4.9$ to $-5.1$  \\ 
1181    &   $\sim0~$ &   $6$      &  3C58           & CCSN  &   2.0--3.2   & 2.1   & $-3.4$ to $-5.4$  \\
1006    &   $-7.5$     &   $36$   & SNR 1006        & SN Ia &   2.1--2.3   & 0.3   & $-7.3$ to $-7.5$  \\
 185    &   $-6.0$     &   $~8+$  & RCW 86          & SN Ia &   2.5--3.2   & 2.5   & $-4.4$ to $-4.9$  \\
1572    &   $-4.5$   &   $18$     &  Tycho's SNR    & SN Ia &   3.0--5.0   & 2.0   & $-4.0$ to $-5.0$  \\
1604    &   $-3.0$     &   $13$   &  Kepler's SNR   & SN Ia &   3.0--5.3   & 2.8   & $-3.2$ to $-4.4$   \\ 
\enddata
\tablenotetext{a}{Estimated peak magnitudes and durations are from 
\citet{CS77}, \citet{Brecher83}, \citet{Schaefer96},  \citet{SG02}, \citet{Ruiz-Lapuente04}, and \citet{Stephenson10}. 
The value for SN 393 reflects the brightness adopted by \citet{CS77} } 
\tablenotetext{b}{Estimates of SN subtype are based on the presence (CCSN) or absence (SN Ia) 
                  of a compact stellar remnant in the confirmed or proposed associated SN remnant. }
\tablenotetext{c}{Estimated peak apparent visual brightnesses were calculated 
                  using the $M_{\rm V}^{\rm max}$ (SN Ia) =  $-19.4$, \citep{Lei01,Riess05}; $M_{\rm V}^{\rm max}$ (CCSN) =  $-18.0$, 
                   \citep{Richardson02,Richardson06}. } 
\tablerefs{{\bf SN 185:} \citet{West69,Rosado96,Vink06,Yam08};
           {\bf SN 393:} \citet{Koyama97,Fukui03,Uch03,Koo03,Koo04,Aharonian04,CC04,Moriguchi05,Tanaka08,Acero09};
           {\bf SN 1006:} \citet{SM80,Winkler03};
           {\bf SN 1054:} \citet{Trimble68,DF85};
           {\bf SN 1181:} \citet{GG82,Roberts93,Fesen08,Kothes10};
           {\bf SN 1572:} \citet{vdb78,Ruiz-Lapuente04,Warren05,Krause08,CC07,Hay10};
           {\bf SN 1604:} \citet{Rey99,Reynolds07}.  }
\label{tab:hist}
\end{deluxetable}

\begin{deluxetable}{lcccc}
\tablecolumns{5}
\tablecaption{Apparent SN 393 Magnitudes by SN Type\tablenotemark{a} }
\tablewidth{0pt}
\tablehead{ \colhead{SN} & \colhead{$M_{\rm V}$ } & 
\multicolumn{3}{c}{\underline{~ ~ ~ ~ ~ ~ ~$m_{\rm v}$\tablenotemark{c} ~ ~ ~ ~ ~ ~ ~ ~ ~}}  \\
\colhead{Type\tablenotemark{b}}   &  \colhead{ }             & \colhead{ ~ ~ ~ t$_{0}$ ~ ~ ~} &  
\colhead{t$_{240}$ ~} & \colhead{t$_{360}$ ~}   }
\startdata
SN~II-P    &  $-17.0$  &  $-3.5$    & $+1.0$        & $+2.0$  \\
SN~II-L    &  $-18.0$  &  $-4.5$    & $+1.5$        & $+2.5$  \\
SN~IIb     &  $-17.5$  &  $-4.0$    & $+1.5$        & $+2.5$  \\
SN~Ib,c    &  $-18.5$  &  $-5.0$    & $+1.0$        & $+3.0$  \\
SN~IIn     &  $-19.0$  &  $-5.5$    & $~0.0$        & $+1.0$  \\
\enddata
\tablenotetext{a}{Rounded to the nearest half magnitude assuming d = 1.3 kpc and $A_{\rm V}$ = 3.0 }
\tablenotetext{b}{Assuming SN light curves from \citet{Turatto90} and templates provided
by P.\ Nugent (\url{http://supernova.lbl.gov/\textasciitilde nugent/nugent\_templates.html}). 
Reported values can deviate $\pm1.0$ mag or more. For SN~IIn, the t$_{360}$ value is an extrapolation  
from the slope of late-time light curves. Peak absolute magnitudes are from \citet{Richardson06}. }
\tablenotetext{c}{Estimated visual magnitudes for March 393, September 393, 
and March 394, i.e., days 0, 240 and 360 post-maximum. }
\label{tab:lc}
\end{deluxetable}

%%%%%%%%%%%%%%%%%%%%%%%%%%%%%%%%%%%%
%%%% Figures
%%%%%%%%%%%%%%%%%%%%%%%%%%%%%%%%%%%%

\end{document}